\documentclass[preprint,preprintnumbers,amsmath,amssymb]{revtex4}
\usepackage{amsfonts}    
\usepackage{amssymb}
\usepackage{latexsym}
\usepackage{eepic}
\usepackage{graphicx}
\usepackage[usenames,dvipsnames]{color}

\newcommand{\al}{\alpha}
\newcommand{\pa}{\partial}
\newcommand{\ep}{\epsilon}

\newcommand{\la}{\lambda}

\newcommand{\Om}{\Omega}
\newcommand{\om}{\omega}

\newcommand{\De}{\Delta}

\newcommand{\tha}{\theta}

\newcommand{\rar}{\rightarrow}

\newcommand{\non}{\nonumber}

\begin{document}

\title{The quantum $H_4$ integrable system}

\author{Marcos A. G. Garc\'ia}
\email{alejandro.garcia@nucleares.unam.mx}
\affiliation{Instituto de Ciencias Nucleares,
      Universidad Nacional Aut\'onoma de M\'exico,
      Apartado Postal 70-543, 04510 M\'exico, D.F., Mexico}

\author{Alexander V. Turbiner}
\email{turbiner@nucleares.unam.mx}
\affiliation{Instituto de Ciencias Nucleares,
    Universidad Nacional Aut\'onoma de M\'exico,
    Apartado Postal 70-543, 04510 M\'exico, D.F., Mexico}

\date{November 9, 2010}

\begin{abstract}
The quantum $H_4$ integrable system is a $4D$ system with rational
potential related to the non-crystallographic root system $H_4$ with
600-cell symmetry. It is shown that the gauge-rotated $H_4$ Hamiltonian
as well as one of the integrals, when written in terms of the
invariants of the Coxeter group $H_4$, is in algebraic form:
it has polynomial coefficients in front of derivatives. Any
eigenfunctions is a polynomial multiplied by ground-state function
(factorization property).
Spectra corresponds to one of the anisotropic harmonic oscillator.
The Hamiltonian has infinitely-many finite-dimensional invariant
subspaces in polynomials, they form the infinite flag with the
characteristic vector $\vec \al\ =\ (1,5,8,12)$.
\end{abstract}

\maketitle

\section{The Hamiltonian}

The quantum $H_4$ system is a four-dimensional system related to the
non-crystallographic root system $H_4$ \cite{Olshanetsky:1983}. The Hamiltonian
of this model is invariant with respect to the $H_4$ Coxeter group, which is
the full symmetry group of the 600-cell polytope. The $H_4$ Coxeter group is
discrete subgroup of $O(4)$ and its dimension is 14400. In Cartesian
coordinates the $H_4$ rational Hamiltonian has the form (see \cite{Olshanetsky:1983})
\begin{equation}
\label{H_H4}
\begin{aligned}
    \mathcal{H}_{H_4} =\ &\frac{1}{2}\sum_{k=1}^{4}\left[-\frac{\partial^{2}}{\partial
    x_k^2}+\omega^{2}x_{k}^{2}+\frac{g}{x_{k}^{2}}\right]+
    \sum_{\mu_{2,3,4}=0,1}\frac{2g}{[x_1+(-1)^{\mu_2}x_2+(-1)^{\mu_3}x_3+(-1)^{\mu_4}x_4]^2}\\
    &+\sum_{\{i,j,k,l\}}\,
    \sum_{\mu_{1,2}=0,1}\frac{2g}{[x_{i}+(-1)^{\mu_1}\varphi_{+}x_{j}+(-1)^{\mu_2}
    \varphi_{-}x_{k}+0\cdot x_{l}]^{2}}\ ,
\end{aligned}
\end{equation}
where $\{i,j,k,l\}=\{1,2,3,4\}$ and its even permutations. Here
$g=\nu(\nu-1)>-1/4$ is the coupling constant,
$\varphi_{\pm}=(1\pm\sqrt{5})/2$ the {\em golden ratio} and its
algebraic conjugate. The configuration space is the subspace of
$\mathbf{R}^4$ where the condition $(\al \cdot x)>0$ holds for
any positive root $\al$ of $H_4$. It is an analogue of
the principal Weyl chamber in the case of crystallographic root
systems.

The ground state eigenfunction and its eigenvalue are
\begin{equation}
\label{Psi_H4}
 \Psi_{0}(x)=\Delta_{1}^{\nu}\Delta_{2}^{\nu}\Delta_{3}^{\nu}
 \exp \bigg(-\frac{\om}{2}\sum_{k=1}^{4}x_{k}^{2} \bigg)\ ,
 \quad E_{0}=2\om(1+30\nu)\ ,
\end{equation}
where
\begin{eqnarray}
\label{D123}
 \De_{1}&=&\prod_{k=1}^{4}x_{k},\\
 \De_{2}&=&\prod_{\mu_{2,3,4}=0,1}[x_1+(-1)^{\mu_2}x_2+(-1)^{\mu_3}x_3+(-1)^{\mu_4}x_4],\\
 \De_{3}&=&\prod_{\{i,j,k,l\}}\,\prod_{\mu_{1,2}=0,1}[x_{i}+(-1)^{\mu_1}\varphi_{+}x_{j}+(-1)^{\mu_2}\varphi_{-}x_{k}+0\cdot
 x_l]\,.
\end{eqnarray}
The ground state eigenfunction (\ref{Psi_H4}) does not vanish in the
configuration space.

The main object of our study is the gauge-rotated Hamiltonian
(\ref{H_H4}) with the ground state eigenfunction (\ref{Psi_H4})
taken as a factor,
\begin{equation}
\label{h_H4}
 h_{\rm H_4} \ =\ -2(\Psi_{0})^{-1}({\cal H}_{\rm H_4}-E_0)(\Psi_{0}) \ ,
\end{equation}
where $E_0$ is the ground state energy given by (\ref{Psi_H4}).
The gauge rotated operator (\ref{h_H4}) is the second-order differential
operator without free term. By construction its lowest eigenfunction is a constant
and the lowest eigenvalue is equal to zero.

Now let us define new variables in (\ref{h_H4}). The $H_4$ root space is characterized
by four fundamental weights $w_c,\ c=1,2,3,4$ (see e.g. \cite{Humphreys:1990}).
Taking action of all group elements on fundamental weight $\om_c$ we generate orbit
$\Om(w_c)$ of a certain length (length $\equiv$ \#elements of the
orbit). The results are summarized as
\begin{center}
\begin{tabular}{cc}
\hline
\rule[-8pt]{0pt}{22pt}{weight}    & \quad orbit length\\
\hline
\rule[-8pt]{0pt}{22pt}{$w_1=(0,\ 0,\ 0,\ 2\varphi_+)$}     & 120 \\
\rule[-8pt]{0pt}{22pt}{$w_2=(1,\ \varphi_+^2,\ 0,\ \varphi_+^4)$} & 600 \\
\rule[-8pt]{0pt}{22pt}{$w_3=(0,\ \varphi_+,\ 1,\ \varphi_+^4-1)$}    & 720 \\
\rule[-8pt]{0pt}{22pt}{$w_4=(0,\ 2\varphi_+,\ 0,\ 2\varphi_+^3)$}    & 1200 \\
\hline\\
\end{tabular}
\end{center}
Now let us find $H_4$-invariants. In order to do it we choose
for simplicity the shortest orbit $\Om (w_1)$ and make averaging,
\begin{equation}
\label{invvars}
    t_{a}^{(\Om)}(x)=\sum_{w \in\ \Om(w_1)}(w \cdot x)^{a}\ ,
\end{equation}
where $a=2,12,20,30$ are the degrees of the $H_4$ group. It is worth noting that
these invariants are defined ambiguously, up to a non-linear combination
of the invariants of the lower degrees
\begin{equation}
\label{coords}
\begin{aligned}
t_{2}^{(\Om)}&\mapsto t_{2}^{(\Om)}\,,\\
t_{12}^{(\Om)}&\mapsto t_{12}^{(\Om)} + A_1 \ (t_{2}^{(\Om)})^6\,,\\
t_{20}^{(\Om)}&\mapsto t_{20}^{(\Om)} + A_2\ (t_{2}^{(\Om)})^{4}t_{12}^{(\Om)} + A_3 \ (t_{2}^{(\Om)})^{10}\,,\\
t_{30}^{(\Om)}&\mapsto t_{30}^{(\Om)} + A_4\
(t_{2}^{(\Om)})^{5}t_{20}^{(\Om)} + A_5 \
(t_{2}^{(\Om)})^{3}(t_{12}^{(\Om)})^{2} + A_6 \
(t_{2}^{(\Om)})^{9}(t_{12}^{(\Om)}) + A_7 \ (t_{2}^{(\Om)})^{15}\,,
\end{aligned}
\end{equation}
where $\{A\}$ are parameters. Canonical invariant basis for the $H_4$ was found only
recently by Mehta \cite{Mehta:1988} (see also \cite{Iwasaki}). Now we use $H_4$ invariants
as new coordinates in (\ref{h_H4}).

Now we can make a change of variables in the gauge-rotated Hamiltonian (\ref{h_H4}):
\[
(x_1,x_2,x_3,x_4) \rar (t_{2}^{(\Om)},t_{12}^{(\Om)},t_{20}^{(\Om)},t_{30}^{(\Om)})\ .
\]
The first observation is that the transformed Hamiltonian $h_{\rm H_4}(t)$ (\ref{h_H4})
takes on an algebraic form for any value of the parameters $\{A\}$ in variables $t$'s (\ref{coords}). The second observation is that for any value of the parameters $\{A\}$
the operator $h_{\rm H_4}(t)$ has infinitely-many finite-dimensional invariant subspaces
in polynomials which form infinite flag. Our goal is to find the parameters for which
$h_{\rm H_4}(t)$ preserves a minimal flag (for a discussion see e.g. \cite{Turbiner:2005_1}).
After some analysis such a set of parameters is found
\begin{equation}
\begin{aligned}
A_1\ =\ -1\,,\quad A_2\ =\ -\frac{43510}{1809}\,,\quad A_3\ =\
\frac{41701}{1809}\,,\quad A_4\ =\
-\frac{17583778485}{146142376}\,,\\
A_5\ =\ -\frac{313009515}{15383408}\,,\quad A_6\ =\
\frac{22081114965}{7691704}\,,\quad A_7\ =\
-\frac{798259915667}{292284752}\,.
\end{aligned}
\end{equation}
Hereafter we call the $t$-variables for such values of parameters as $\tau$-variables.

In order to write down explicit expressions for variables $\tau$ it is useful to exploit the notation for multivariate polynomials introduced by Iwasaki et al \cite{Iwasaki}.
For given partition $\la\ =\ (\la_1,\la_2,\la_3,\la_4)$ with $\la_1\geq\la_2\geq\la_3\geq\la_4\geq0$, let denote as $M_{\la}$ the associated
monomial symmetric polynomial of the variables $(x_1^2,x_2^2,x_3^2,x_4^2)$,
\[
M_{\la}=\sum x_1^{2\mu_1} x_2^{2\mu_2} x_3^{2\mu_3} x_4^{2\mu_4}\ ,
\]
where the sum is taken over all permutations
$(\mu_1,\mu_2,\mu_3,\mu_4)$ of
$(\lambda_1,\lambda_2,\lambda_3,\lambda_4)$. If $\lambda$ consists
of mutually distinct numbers $p_1>\cdots>p_m$ with $p_j$ appearing
$k_j$ times in $\lambda$, then we denote the polynomial as
\[
M_{\lambda}=[p_1^{k_1}|\cdots|p_m^{k_m}]\ .
\]
Let introduce also $\De_4$ as the fundamental alternating polynomial
of $(x_1^2,x_2^2,x_3^2,x_4^2)$,
\[
\Delta_4=\prod_{1\leq i<j\leq 4}(x_i^2-x_j^2)\ .
\]
Then, the $\tau$-variables (7)-(9)are written in these notations as
\begin{eqnarray}
\label{tau1}
\tau_1&=& [1|0^3]\equiv x_1^2+x_2^2+x_3^2+x_4^2\ ,\non\\[10pt]
\tau_2&=& 14[3^2|0^2]-6[4|2|0^2]+2[5|1|0^2]-270[2^2|1^2]+30[2^3|0]\non\\
&&-12[4|1^2|0]+348[3|1^3]+9[3|2|1|0]+33\sqrt{5}\De_4\ ,\non\\[10pt]
\tau_3&=& 2[8|2|0^2]+4[8|1^2|0]-10[7|3|0^2]-45[7|2|1|0]+60[7|1^3]\non\\
&&+22[6|4|0^2]+157[6|3|1|0]+270[6|2^2|0]-150[6|2|1^2]\non \\
&&-22[5^2|0^2]-131[5|4|1|0]-733[5|3|2|0]-2156[5|3|1^2]\non\\
&&+4050[5|2^2|1]+1320[4^2|2|0]+4650[4^2|1^2]+6[4|3^2|0]\non\\
&&-2175[4|3|2|1]-19050[4|2^3]+10800[3^2|2^2]+3336[3^3|1]\non\\
&&+3\sqrt{5}\De_4\{\ 5[4|0^3]-18[3|1|0^2]+49[2^2|0^2]+3[2|1^2|0]\non\\
&&+1146[1^4]\ \}\ ,\non
\end{eqnarray}
\begin{eqnarray}
\label{tau2}
\tau_4&=&\ 65742[15|0^3]-504[13|2|0^2]+830[13|1^2|0]+61690[12|3|0^2]\non\\
&&-5130[12|2|1|0]-9495[12|1^3]+18795[11|4|0^2]\non\\
&&+28560[11|3|1|0]-43500[11|2^2|0]-53070[11|2|1^2]\non\\
&&-156330[10|5|0^2]+59130[10|4|1|0]+26415[10|3|2|0]\non\\
&&+405255[10|3|1^2]+1350[10|2^2|1]+19710[9|6|0^2]\non\\
&&-20[9|4|2|0]-8663355[9|4|1^2]-120[9|3^2|0]+450[9|3|2|1]\non\\
&&-962715[9|2^3]+13860[8|7|0^2]-94530[8|6|1|0]\non\\
&&-353160[8|5|2|0]-1452060[8|5|1^2]+5557050[8|4|3|0]\non\\
&&+590580[8|4|2|1]-198270[8|3^2|1]+389250[7^2|1|0]\non\\
&&+2897820[7|6|2|0]-5227920[7|6|1^2]+1134540[7|5|3|0]\\
&&-4041270[7|5|2|1]-591330[7|4^2|0]+23417850[7|4|3|1]\non\\
&&-22770[7|4|2^2]-23528790[7|3^2|2]+29647380[6^2|3|0]\non\\
&&+36597510[6^2|2|1]-1649925[6|5|4|0]+150[6|5|3|1]\non\\
&&+40935[6|5|2^2]-510[6|4^2|1]-60[6|4|3|2]+242505[6|3^3]\non\\
&&+270060[5^3|0]-528270[5^2|4|1]-36255[5|4^2|2]+825[5|4|3^2]\non\\
&&+707085[4^3|3]+45\sqrt{5}\De_4\{\ -27040[9|0^3]-5[8|1|0^2]\non\\
&&-1914[7|1^2|0]+23[6|3|0^2]+91[6|2|1|0]-44[6|1^3]\non\\
&&-352[5|4|0^2]+8[5|3|1|0]+1085[5|2^2|0]+6875[5|2|1^2]\non\\
&&-5168[4^2|1|0]-934[4|3|2|0]-568[4|3|1^2]+1773[4|2^2|1]\non\\
&&+20911[3^3|0]+15915[3^2|2|1]+573[3|2^3]\ \}\ .\non
\end{eqnarray}
Thus, the variables $\tau_{1,2,3,4}$ are homogeneous polynomials in $x^2$'s
of the degrees 1,3,10,15 , respectively.

Finally, the gauge-rotated Hamiltonian (\ref{h_H4}) in the
$\tau$-coordinates (\ref{tau2}) written as
\begin{equation}
\label{h_H4_tau}
 {h}_{\rm H_4}\ =\
 \sum_{i,j=1}^{4} {A}_{ij}({\tau})
 \frac{\pa^2}{\pa {{\tau}_i} \pa {{\tau}_j} } +
 \sum_{i=1}^{4}{B}_i({\tau}) \frac{\pa}{\pa {\tau}_i} \quad ,
 \quad {A}_{ij}={A}_{ji} \ ,
\end{equation}
takes amazingly simple form with the coefficient functions
\[
A_{11}\ =\ 4 \ \tau_1 \ ,\ A_{12}\ =\ 24\ \tau_2 \ ,
\]
\[
A_{13}\ =\ 40\ \tau_3 \ ,\ A_{14}\ =\ 60\ \tau_4 \ ,
\]
\begin{eqnarray}
\label{A-B}
A_{22}&=& 88\ \tau_1\tau_3 + 8\ \tau_1^5\tau_2 \ ,\non \\
A_{23}&=& -4\ \tau_1^3\tau_2^2 + 24\ \tau_1^5\tau_3 - 8\ \tau_4 \ ,\non\\
A_{24}&=& 10\ \tau_1^2\tau_2^3 + 60\ \tau_1^4\tau_2\tau_3 + 40\ \tau_1^5\tau_4
- 600\ \tau_3^2 \ ,\non\\
A_{33}&=& -\frac{38}{3}\ \tau_1\tau_2^3 + 28\ \tau_1^3\tau_2\tau_3 - \frac{8}{3}\ \tau_1^4\tau_4 \ ,\non \\
A_{34}&=& 210\ \tau_1^2\tau_2^2\tau_3 + 60\ \tau_1^3\tau_2\tau_4
- 180\ \tau_1^4\tau_3^2 + 30\ \tau_2^4 \ ,\non \\
A_{44}&=& -2175\ \tau_1\tau_2^3\tau_3 - 450\tau_1^2\tau_2^2\tau_4 - 1350\ \tau_1^3\tau_2\tau_3^2 - 600\ \tau_1^4\tau_3\tau_4 \ ,
\end{eqnarray}
\begin{eqnarray}
B_{1}&=&  8(1+30 \nu)\ - 4\om\tau_1 \ ,\non \\
B_{2}&=& 12(1+10 \nu)\ \tau_1^5\ -\ 24\om\tau_2 \ ,\non \\
B_{3}&=& 20(1+6  \nu)\ \tau_1^3\tau_2\ -\ 40\om\tau_3 \ ,\non \\
B_{4}&=& 15(1-30 \nu)\ \tau_1^2\tau_2^2\ -\ 450(1+2\nu)\
\tau_1^4\tau_3\ -\ 60\om\tau_4 \ . \non
\end{eqnarray}

It can be easily checked that the operator (\ref{h_H4_tau}) is
triangular with respect to action on monomials
$\tau_1^{p_1}\tau_2^{p_2}\tau_3^{p_3}\tau_4^{p_4}$. One can find the
spectrum of (\ref{h_H4_tau}) $h_{\rm H_4} \varphi=-2\epsilon\varphi$
explicitly
\begin{equation}
\label{spectrumh}
\ep_{n_{1},n_{2},n_{3},n_{4}} = 2\om(n_{1}+6n_{2}+10n_{3}+15n_4)\,,
\end{equation}
where $n_i=0,1,2,\ldots$. Degeneracy of the spectrum is related to
the number of solutions of the equation $n_1+6n_2+10n_3+15n_4=n$ for
$n=0,1,2\ldots$ in non-negative numbers $n_{1,2,3,4}$. The spectrum
$\ep$ does not depend on the coupling constant $g$ and it is
equidistant. It coincides to the spectrum of $4D$ anisotropic
harmonic oscillator with frequencies $(2\om,12\om,20\om,30\om)$. The
energies of the original rational $H_4$ Hamiltonian (\ref{H_H4}) are
$E=E_0+\ep$. It is worth noting that the Hamiltonian (\ref{h_H4_tau})
has infinite family of eigenfunctions $\phi_{n_1,0,0,0}$ depending
on single variable $\tau_1$. They are given by the Laguerre polynomials
and the eigenvalues are linear in quantum number (cf.(\ref{spectrumh}))
\begin{equation}
\label{laguerre}
\phi_{n_1,0,0,0}(\tau_1)\ =\ L_{n_1}^{(1+60\nu)}(\om\tau_1)\ ,\quad
\ep_{n_1,0,0,0}=2\om n_1\ ,\quad n_1=0,1,2,\ldots\ .
\end{equation}

The boundary of the configuration space of the rational $H_4$ model
(\ref{H_H4}) in the $\tau$ variables is determined by the zeros of
the ground state eigenfunction, hence, by pre-exponential factor in
(\ref{Psi_H4}). It is the algebraic surface of degree 120 in
Cartesian coordinates being a product of monomials. In
$\tau$-coordinates (\ref{tau2}) it can be written as

\begin{equation*}
\begin{aligned}
&64\ \tau_1^{15}\tau_4^3 + 1440\ \tau_1^{14}\tau_2\tau_3\tau_4^2 +
 10800\ \tau_1^{13}\tau_2^2\tau_3^2\tau_4 + 27000\ \tau_1^{12}\tau_2^3\tau_3^3 -
 240\ \tau_1^{12}\tau_2^3\tau_4^2 \\
&- 3600\ \tau_1^{11}\tau_2^4\tau_3\tau_4 - 13500\ \tau_1^{10}\tau_2^5\tau_3^2 +
 34992\ \tau_1^{10}\tau_3^5 - 1440\ \tau_1^{10}\tau_3^2\tau_4^2 +
 300\ \tau_1^9\tau_2^6\tau_4 \\
&- 2160\ \tau_1^9\tau_2\tau_3^3\tau_4 - 1440\tau_1^9\tau_2\tau_4^3 +
 2250\ \tau_1^8\tau_2^7\tau_3 - 22680\ \tau_1^8\tau_2^2\tau_3^4 -
 28080\ \tau_1^8\tau_2^2\tau_3\tau_4^2 \\
&- 203760\ \tau_1^7\tau_2^3\tau_3^2\tau_4 - 125\ \tau_1^6\tau_2^9 -
 493020\ \tau_1^6\tau_2^4\tau_3^3 + 3600\ \tau_1^6\tau_2^4\tau_4^2 +
 57780\ \tau_1^5\tau_2^5\tau_3\tau_4\\
&- 8640\ \tau_1^5\tau_3^4\tau_4 + 4320\ \tau_1^5\tau_3\tau_4^3 +
 221310\ \tau_1^4\tau_2^6\tau_3^2 -
 648000\ \tau_1^4\tau_2\tau_3^5 + 116640\ \tau_1^4\tau_2\tau_3^2\tau_4^2 \\
&- 4680\ \tau_1^3\tau_2^7\tau_4 + 712800\ \tau_1^3\tau_2^2\tau_3^3\tau_4 +
 6480\ \tau_1^3\tau_2^2\tau_4^3 - 35640\ \tau_1^2\tau_2^8\tau_3
 + 2052000\ \tau_1^2\tau_2^3\tau_3^4
\end{aligned}
\end{equation*}
\begin{equation}
\label{confspacet}
\begin{aligned}
&+ 62640\ \tau_1^2\tau_2^3\tau_3\tau_4^2 + 259200\ \tau_1\tau_2^4\tau_3^2\tau_4 +
  1944\ \tau_2^{10} + 129600\ \tau_2^5\tau_3^3 + 2592\ \tau_2^5\tau_4^2 \\
&+ 2160000\ \tau_3^6 - 86400\ \tau_3^3\tau_4^2 + 864\ \tau_4^4\ =\ 0\ ,
\end{aligned}
\end{equation}
which is the algebraic surface of degree 18 being given by a polynomial of degree 15 in $\tau_1$, of the degree 10 in $\tau_2$, of the degree 6 in $\tau_3$ and of the degree 4 in $\tau_4$. It is worth mentioning that l.h.s. of (\ref{confspacet}) is proportional to the square of Jacobian, $ J^2 (\frac{\pa \tau}{\pa x}) $.

The Hamiltonian $h_{\rm H_4}(\tau)$ has infinitely-many finite-dimensional invariant subspaces
\begin{equation}
\label{minflag}
 \mathcal{P}_{n}^{(1,5,8,12)}=\langle
 \tau_1^{p_1}\tau_2^{p_2}\tau_3^{p_3}\tau_4^{p_4}\,|
 \,0\leq p_1+5p_2+8p_3+12p_4\leq n\rangle\ , \quad n=0,1,2,\ldots \ ,
\end{equation}
which form the (minimal) infinite flag. Its characteristic vector is
\begin{equation}
\label{min-ch-vec}
   \vec{\al}_{min}\ =\ (1,5,8,12)\ .
\end{equation}
It is worth noting that each particular space
$\mathcal{P}_{n}^{(1,5,8,12)}$ (\ref{minflag}) as well as the whole
flag are invariant with respect to a weighted projective
transformation
\begin{eqnarray}
\label{wpt}
\tau_1 &\rightarrow& \tau_1+a\,,\non\\
\tau_2 &\rightarrow& \tau_2+b_1\,\tau_1^5+b_2\,\tau_1^4+b_3\,\tau_1^3+b_4\,\tau_1^2+b_5\,\tau_1+b_6\,,\non\\
\tau_3 &\rightarrow& \tau_3+c_1\,\tau_1^3\tau_2+c_2\,\tau_1^2\tau_2+c_3\,\tau_1\tau_2+c_4\,\tau_2+c_5\,\tau_1^8+c_6\,\tau_1^7\non\\
&&\ +c_7\,\tau_1^6+c_8\,\tau_1^5+c_9\,\tau_1^4+c_{10}\,\tau_1^3+c_{11}\,\tau_1^2+c_{12}\,\tau_1+c_{13}\,,\\
\tau_4 &\rightarrow& \tau_4+d_1\,\tau_1^4\tau_3+d_2\,\tau_1^3\tau_3+d_3\,\tau_1^2\tau_3+d_4\,\tau_1\tau_3+d_5\,\tau_3\non\\
&&\ +d_6\,\tau_1^7\tau_2+d_7\,\tau_1^6\tau_2+d_8\,\tau_1^5\tau_2+d_9\,\tau_1^4\tau_2+d_{10}\,\tau_1^3\tau_2\non\\
&&\ +d_{11}\,\tau_1^2\tau_2+d_{12}\,\tau_1\tau_2+d_{13}\,\tau_2+d_{14}\,\tau_1^{12}+d_{15}\,\tau_1^{11}\non\\
&&\ +d_{16}\,\tau_1^{10}+d_{17}\,\tau_1^9+d_{18}\,\tau_1^8+d_{19}\,\tau_1^7+d_{20}\,\tau_1^6+d_{21}\,\tau_1^5\non\\
&&\
+d_{22}\,\tau_1^4+d_{23}\,\tau_1^3+d_{24}\,\tau_1^2+d_{25}\,\tau_1+d_{26}\,,\non
\end{eqnarray}
where $\{a,b,c,d\}$ are parameters. It manifests a hidden invariance
of the Hamiltonian (\ref{H_H4}) preserving its algebraic form. A meaning of this
invariance is unclear.

\medskip

\section{Integral}

The Hamiltonian (\ref{H_H4}) can be written in {\em hyperspherical
coordinates}
\begin{equation}
\label{hyps_coords}
\begin{aligned}
x_1=\ &r\sin\psi\sin\theta\cos\phi\ ,\\
x_2=\ &r\sin\psi\sin\theta\sin\phi\ ,\\
x_3=\ &r\sin\psi\cos\theta\ ,\\
x_4=\ &r\cos\psi\ ,
\end{aligned}
\end{equation}
where it takes the form
\begin{equation}
\label{H_H_4_sphe}
{\cal H}_{H_4}=-\frac{1}{2}\De^{(4)} +
\frac{1}{2}\om^2 r^2 + \frac{W(\psi,\theta,\phi)}{r^2}\ .
\end{equation}
Here $\De^{(4)}$ is the $4D$ Laplacian and the angular function

\begin{equation*}
\begin{aligned}
W(\psi,\theta,\phi)= &\
\frac{2\nu(\nu-1)}{(s_{\psi}s_{\tha}c_{\phi}+\varphi_+s_{\psi}s_{\tha}s_{\phi}+\varphi_
-s_{\psi}c_{\tha})^2}
+\frac{2\nu(\nu-1)}{(s_{\psi}s_{\tha}c_{\phi}-\varphi_+s_{\psi}s_{\tha}s_{\phi}+\varphi_
-s_{\psi}c_{\tha})^2}\\
&+\frac{2\nu(\nu-1)}{(s_{\psi}s_{\tha}c_{\phi}+\varphi_+s_{\psi}s_{\tha}s_{\phi}-\varphi_
-s_{\psi}c_{\tha})^2}
+\frac{2\nu(\nu-1)}{(s_{\psi}s_{\tha}c_{\phi}-\varphi_+s_{\psi}s_{\tha}s_{\phi}-\varphi_
-s_{\psi}c_{\tha})^2}\\
&+\frac{2\nu(\nu-1)}{(s_{\psi}s_{\tha}c_{\phi}+\varphi_+s_{\psi}c_{\tha}+\varphi_-c_{\psi})^2}
+\frac{2\nu(\nu-1)}{(s_{\psi}s_{\tha}c_{\phi}-\varphi_+s_{\psi}c_{\tha}+\varphi_-c_{\psi})^2}\\
&+\frac{2\nu(\nu-1)}{(s_{\psi}s_{\tha}c_{\phi}+\varphi_+s_{\psi}c_{\tha}-\varphi_-c_{\psi})^2}
+\frac{2\nu(\nu-1)}{(s_{\psi}s_{\tha}c_{\phi}-\varphi_+s_{\psi}c_{\tha}-\varphi_-c_{\psi})^2}\\
&+\frac{2\nu(\nu-1)}{(s_{\psi}s_{\tha}c_{\phi}+\varphi_+c_{\psi}+\varphi_-s_{\psi}s_{\tha}s_{\phi})^2}
+\frac{2\nu(\nu-1)}{(s_{\psi}s_{\tha}c_{\phi}-\varphi_-c_{\psi}+\varphi_-s_{\psi}s_{\tha}s_{\phi})^2}\\
&+\frac{2\nu(\nu-1)}{(s_{\psi}s_{\tha}c_{\phi}+\varphi_+c_{\psi}-\varphi_-s_{\psi}s_{\tha}s_{\phi})^2}
+\frac{2\nu(\nu-1)}{(s_{\psi}s_{\tha}c_{\phi}-\varphi_+c_{\psi}-\varphi_-s_{\psi}s_{\tha}s_{\phi})^2}\\
&+\frac{2\nu(\nu-1)}{(s_{\psi}s_{\tha}s_{\phi}+\varphi_+s_{\psi}s_{\tha}c_{\phi}+\varphi_-c_{\psi})^2}
+\frac{2\nu(\nu-1)}{(s_{\psi}s_{\tha}s_{\phi}-\varphi_+s_{\psi}s_{\tha}c_{\phi}+\varphi_-c_{\psi})^2}\\
&+\frac{2\nu(\nu-1)}{(s_{\psi}s_{\tha}s_{\phi}+\varphi_+s_{\psi}s_{\tha}c_{\phi}-\varphi_-c_{\psi})^2}
+\frac{2\nu(\nu-1)}{(s_{\psi}s_{\tha}s_{\phi}-\varphi_+s_{\psi}s_{\tha}c_{\phi}-\varphi_-c_{\psi})^2}\\
&+\frac{2\nu(\nu-1)}{(s_{\psi}s_{\tha}s_{\phi}+\varphi_+s_{\psi}c_{\tha}+\varphi_-s_{\psi}s_{\tha}c_{\phi})^2}
+\frac{2\nu(\nu-1)}{(s_{\psi}s_{\tha}s_{\phi}-\varphi_+s_{\psi}c_{\tha}+\varphi_-s_{\psi}s_{\tha}c_{\phi})^2}\\
&+\frac{2\nu(\nu-1)}{(s_{\psi}s_{\tha}s_{\phi}+\varphi_+s_{\psi}c_{\tha}-\varphi_-s_{\psi}s_{\tha}c_{\phi})^2}
+\frac{2\nu(\nu-1)}{(s_{\psi}s_{\tha}s_{\phi}-\varphi_-s_{\psi}c_{\tha}-\varphi_-s_{\psi}s_{\tha}c_{\phi})^2}\\
\end{aligned}
\end{equation*}
\begin{equation}
\begin{aligned}
&+\frac{2\nu(\nu-1)}{(s_{\psi}s_{\tha}s_{\phi}+\varphi_+c_{\psi}+\varphi_-s_{\psi}c_{\tha})^2}
+\frac{2\nu(\nu-1)}{(s_{\psi}s_{\tha}s_{\phi}-\varphi_+c_{\psi}+\varphi_-s_{\psi}c_{\tha})^2}\\
&+\frac{2\nu(\nu-1)}{(s_{\psi}s_{\tha}s_{\phi}+\varphi_+c_{\psi}-\varphi_-s_{\psi}c_{\tha})^2}
+\frac{2\nu(\nu-1)}{(s_{\psi}s_{\tha}s_{\phi}-\varphi_+c_{\psi}-\varphi_-s_{\psi}c_{\tha})^2}\\
&+\frac{2\nu(\nu-1)}{(s_{\psi}c_{\tha}+\varphi_+s_{\psi}s_{\tha}c_{\phi}+\varphi_-s_{\psi}s_{\tha}s_{\phi})^2}
+\frac{2\nu(\nu-1)}{(s_{\psi}c_{\tha}+\varphi_-s_{\psi}s_{\tha}c_{\phi}+\varphi_-s_{\psi}s_{\tha}s_{\phi})^2}\\
&+\frac{2\nu(\nu-1)}{(s_{\psi}c_{\tha}+\varphi_+s_{\psi}s_{\tha}c_{\phi}-\varphi_-s_{\psi}s_{\tha}s_{\phi})^2}
+\frac{2\nu(\nu-1)}{(s_{\psi}c_{\tha}+\varphi_-s_{\psi}s_{\tha}c_{\phi}-\varphi_-s_{\psi}s_{\tha}s_{\phi})^2}\\
&+\frac{2\nu(\nu-1)}{(s_{\psi}c_{\tha}+\varphi_+c_{\psi}+\varphi_-s_{\psi}s_{\tha}c_{\phi})^2}
+\frac{2\nu(\nu-1)}{(s_{\psi}c_{\tha}-\varphi_+c_{\psi}+\varphi_-s_{\psi}s_{\tha}c_{\phi})^2}\\
&+\frac{2\nu(\nu-1)}{(s_{\psi}c_{\tha}+\varphi_+c_{\psi}-\varphi_-s_{\psi}s_{\tha}c_{\phi})^2}
+\frac{2\nu(\nu-1)}{(s_{\psi}c_{\tha}-\varphi_+c_{\psi}-\varphi_-s_{\psi}s_{\tha}c_{\phi})^2}\\
&+\frac{2\nu(\nu-1)}{(s_{\psi}c_{\tha}+\varphi_+s_{\psi}s_{\tha}s_{\phi}+\varphi_-c_{\psi})^2}
+\frac{2\nu(\nu-1)}{(s_{\psi}c_{\tha}+\varphi_-s_{\psi}s_{\tha}s_{\phi}+\varphi_-c_{\psi})^2}\\
&+\frac{2\nu(\nu-1)}{(s_{\psi}c_{\tha}+\varphi_+s_{\psi}s_{\tha}s_{\phi}-\varphi_-c_{\psi})^2}
+\frac{2\nu(\nu-1)}{(s_{\psi}c_{\tha}+\varphi_-s_{\psi}s_{\tha}s_{\phi}-\varphi_-c_{\psi})^2}\\
&+\frac{2\nu(\nu-1)}{(c_{\psi}+\varphi_+s_{\psi}s_{\tha}c_{\phi}+\varphi_-s_{\psi}c_{\tha})^2}
+\frac{2\nu(\nu-1)}{(c_{\psi}+\varphi_-s_{\psi}s_{\tha}c_{\phi}+\varphi_-s_{\psi}c_{\tha})^2}\\
&+\frac{2\nu(\nu-1)}{(c_{\psi}+\varphi_+s_{\psi}s_{\tha}c_{\phi}-\varphi_-s_{\psi}c_{\tha})^2}
+\frac{2\nu(\nu-1)}{(c_{\psi}+\varphi_-s_{\psi}s_{\tha}c_{\phi}-\varphi_-s_{\psi}c_{\tha})^2}\\
&+\frac{2\nu(\nu-1)}{(c_{\psi}+\varphi_+s_{\psi}s_{\tha}s_{\phi}+\varphi_-s_{\psi}s_{\tha}c_{\phi})^2}
+\frac{2\nu(\nu-1)}{(c_{\psi}-\varphi_+s_{\psi}s_{\tha}s_{\phi}+\varphi_-s_{\psi}s_{\tha}c_{\phi})^2}\\
&+\frac{2\nu(\nu-1)}{(c_{\psi}+\varphi_+s_{\psi}s_{\tha}s_{\phi}-\varphi_-s_{\psi}s_{\tha}c_{\phi})^2}
+\frac{2\nu(\nu-1)}{(c_{\psi}-\varphi_+s_{\psi}s_{\tha}s_{\phi}-\varphi_-s_{\psi}s_{\tha}c_{\phi})^2}\\
&+\frac{2\nu(\nu-1)}{(c_{\psi}+\varphi_+s_{\psi}c_{\tha}+\varphi_-s_{\psi}s_{\tha}s_{\phi})^2}
+\frac{2\nu(\nu-1)}{(c_{\psi}-\varphi_+s_{\psi}c_{\tha}+\varphi_-s_{\psi}s_{\tha}s_{\phi})^2}\\
&+\frac{2\nu(\nu-1)}{(c_{\psi}+\varphi_+s_{\psi}c_{\tha}-\varphi_-s_{\psi}s_{\tha}s_{\phi})^2}
+\frac{2\nu(\nu-1)}{(c_{\psi}-\varphi_+s_{\psi}c_{\tha}-\varphi_-s_{\psi}s_{\tha}s_{\phi})^2}\\
&+\frac{2\nu(\nu-1)}{(s_{\psi}s_{\tha}c_{\phi}+s_{\psi}s_{\tha}s_{\phi}+s_{\psi}c_{\tha}+c_{\psi})^2}
+\frac{2\nu(\nu-1)}{(s_{\psi}s_{\tha}c_{\phi}-s_{\psi}s_{\tha}s_{\phi}+s_{\psi}c_{\tha}+c_{\psi})^2}\\
&+\frac{2\nu(\nu-1)}{(s_{\psi}s_{\tha}c_{\phi}+s_{\psi}s_{\tha}s_{\phi}-s_{\psi}c_{\tha}+c_{\psi})^2}
+\frac{2\nu(\nu-1)}{(s_{\psi}s_{\tha}c_{\phi}+s_{\psi}s_{\tha}s_{\phi}+s_{\psi}c_{\tha}-c_{\psi})^2}\\
&+\frac{2\nu(\nu-1)}{(s_{\psi}s_{\tha}c_{\phi}-s_{\psi}s_{\tha}s_{\phi}-s_{\psi}c_{\tha}+c_{\psi})^2}
+\frac{2\nu(\nu-1)}{(s_{\psi}s_{\tha}c_{\phi}-s_{\psi}s_{\tha}s_{\phi}+s_{\psi}c_{\tha}-c_{\psi})^2}\\
&+\frac{2\nu(\nu-1)}{(s_{\psi}s_{\tha}c_{\phi}+s_{\psi}s_{\tha}s_{\phi}-s_{\psi}c_{\tha}-c_{\psi})^2}
+\frac{2\nu(\nu-1)}{(s_{\psi}s_{\tha}c_{\phi}-s_{\psi}s_{\tha}s_{\phi}-s_{\psi}c_{\tha}-c_{\psi})^2}\\
&+\frac{\nu(\nu+1)}{2s_{\psi}^2s_{\tha}^2c_{\phi}^2}
+\frac{\nu(\nu+1)}{2s_{\psi}^2s_{\tha}^2s_{\phi}^2}
+\frac{\nu(\nu+1)}{2s_{\psi}^2c_{\tha}^2}
+\frac{\nu(\nu+1)}{2c_{\psi}^2} \ .
\end{aligned}
\end{equation}
Here, for the sake of simplicity we denoted $c_{\psi}\equiv\cos
\psi$, $s_{\psi}\equiv\sin \psi$, $c_{\tha}\equiv\cos
\tha$, $s_{\tha}\equiv\sin \tha$ and $c_{\phi}\equiv\cos
\phi$, $s_{\phi}\equiv\sin \phi$. It is seen
immediately, that the Schroedinger equation (\ref{H_H_4_sphe})
admits a separation of radial variable $r$: any solution can be
written in factorized form
\begin{equation}
\Psi(r,\psi,\theta,\phi)=R(r)Q(\psi,\theta,\phi)\ .
\end{equation}
Functions $R$ and $Q$ are the solutions of the
equations
\begin{equation}
\bigg[-\frac{1}{2 r^3}\frac{\pa}{\pa r}\bigg(r^3\frac{\pa}{\pa
r}\bigg) + \frac{1}{2}\om^2r^2 + \frac{\gamma}{r^2}\bigg]R(r) \ =\ E
R(r)\ ,
\end{equation}
\begin{equation}
\label{eigf} \mathcal{F}\ Q(\psi,\theta,\phi)=\gamma\
Q(\psi,\theta,\phi)\ ,
\end{equation}
respectively, while $\gamma$ is the constant of separation. The operator
$\mathcal{F}$ has the form
\begin{equation}
\label{opf}
   \mathcal{F}\ =\
    -\frac{1}{2\sin^2\psi}\bigg[\frac{\pa}{\pa\psi}\bigg(\sin^2\psi\frac{\pa}{\pa\psi}\bigg)
    +\frac{1}{\sin\theta}\frac{\pa}{\pa\theta}\bigg(\sin\theta\frac{\pa}{\pa\theta}\bigg)
    + \frac{1}{\sin^2\theta}\frac{\pa^2}{\pa\phi^2}\bigg]
    + W(\psi,\theta,\phi) \ ,
\end{equation}
It can be immediately checked that the Hamiltonian ${\cal H}_{H_4}$
and $\mathcal{F}$ commute,
\begin{equation}
         [{\cal H}_{H_4},\mathcal{F}]\ =\ 0\ .
\end{equation}
Hence, $\mathcal{F}$ is an integral of motion. Thus, it has common
eigenfunctions with the Hamiltonian ${\cal H}_{H_4}$.

Let us make a gauge rotation of the operator $\mathcal{F}$
(\ref{opf}) with the ground state function $\Psi_0$ as a gauge factor,
\begin{equation}
\label{f}
        f\ =\ (\Psi_0)^{-1}(\mathcal{F}-\gamma_{0})\Psi_0\ ,\qquad
\gamma_0=60\nu(1+30\nu)\ ,
\end{equation}
where $\gamma_0$ is the lowest eigenvalue of $\mathcal{F}$. Then make
a change of variables to the $\tau$-variables (\ref{tau2}). The operator
$f$ gets an algebraic form,
\begin{equation}
\label{fmin} f\ =\
\sum_{i,j=1}^{4}F_{ij}\frac{\pa^{2}}{\pa\tau_{i}\pa\tau_{j}}
 +\sum_{j=1}^{4}G_{j}\frac{\pa}{\pa\tau_{j}} \ ,\quad {F}_{ij}={F}_{ji}
\end{equation}
where
\[
F_{11}\ =\ 0\,,\ F_{12}\ =\ 0\,,
\]
\[
F_{13}\ =\ 0\,,\ F_{14}\ =\ 0\,,
\]
\begin{eqnarray}
F_{22}&=&-4\tau_1^6\tau_2-44\tau_1^2\tau_3+72\tau_2^2\,,\non \\
F_{23}&=&-12\tau_1^6\tau_3+2\tau_1^4\tau_2^2+4\tau_1\tau_4+120\tau_2\tau_3\,, \non\\
F_{24}&=&-20\tau_1^6\tau_4-30\tau_1^5\tau_2\tau_3
 -5\tau_1^3\tau_2^3+300\tau_1\tau_3^2+180\tau_2\tau_4\,, \non
\end{eqnarray}
\begin{eqnarray}
F_{33}&=&\frac{4}{3}\tau_1^5\tau_4-14\tau_1^4\tau_2\tau_3
 +\frac{19}{3}\tau_1^2\tau_2^3+200\tau_3^2\,,\non\\
F_{34}&=&90\tau_1^5\tau_3^2-30\tau_1^4\tau_2\tau_4
 -105\tau_1^3\tau_2^2\tau_3-15\tau_1\tau_2^4+300\tau_3\tau_4\,,\non\\
F_{44}&=&300\tau_1^5\tau_3\tau_4+675\tau_1^4\tau_2\tau_3^2
 +225\tau_1^3\tau_2^2\tau_4+\frac{2175}{2}\tau_1^2\tau_2^3\tau_3+450\tau_4^2\,,\non\\
G_{1}&=&0\,,\non\\
G_{2}&=&-6(1+10\nu)\tau_1^6+12(7+60\nu)\tau_2\,,\non \\
G_{3}&=&-10(1+6\nu)\tau_1^4\tau_2+20(11+60\nu)\tau_3\,,\non\\
G_{4}&=&225(1+2\nu)\tau_1^5\tau_3-\frac{15}{2}(1-30\nu)\tau_1^3\tau_2^2+40(12+45\nu)\tau_4\,.
\end{eqnarray}
It is worth noting that in the operator $f$ the variable $\tau_1$
appears as a parameter. It implies that any eigenfunction of the Hamiltonian
$h_{H_4}$, which depends on $\tau_1$ only, is an eigenfunction of the
integral $f$ with zero eigenvalue.

It can be also shown that the operator $f$ has infinitely many finite-dimensional invariant subspaces in polynomials
\begin{equation}
 \mathcal{P}_{n}^{(1,6,10,15)}\ =\
 \langle \tau_1^{p_1}\tau_2^{p_2}\tau_3^{p_3}\tau_4^{p_4}\,|
 \,0\leq p_1+6p_2+10p_3+15p_4\leq n\rangle\ ,\ n=0,1,2, \ldots \ ,
\end{equation}
which form a flag with characteristic vector $(1,6,10,15)$.
The spectrum of the integral $\mathcal{F} \Psi = \Gamma \Psi$ can be found in a closed
form,
\[
\Gamma_{0, k_2, k_3, k_4}\ \equiv\ \gamma_{0, k_2, k_3, k_4} + \gamma_0\ =
\]
\begin{equation}
\label{gamma}
 72k_2^2+200k_3^2+450k_4^2+120k_2k_3+180k_2k_4+300k_3k_4\\
 +2(1+60\nu)(6k_2+10k_3+15k_4)+\gamma_0\ ,
\end{equation}
where $k_2,k_3,k_4=0,1,2,\ldots$ and $\gamma_0$ is given by (\ref{f}).

It can be shown that the Hamiltonian $h_{H_4}$ has a certain
degeneracy -- it preserves two different flags: one with (minimal)
characteristic vector (1,5,8,12) and another one with characteristic
vector (1,6,10,15). The fact that the operator $h_{H_4}$ with
coefficients (\ref{A-B}) commutes with $f$ given by (\ref{fmin})
implies that common eigenfunctions of the operators $h_{H_4}$ and
$f$ are elements of the flag of spaces $\mathcal{P}^{(1,6,10,15)}$.

Let us denote $\phi_{n,i}$ the eigenfunctions of $h_{H_4}$ which are
elements of the invariant space $P^{(1,5,8,12)}_n$ and their respectful
eigenvalues $\ep_{n,i}$. The index $i$ numerates these eigenfunctions for given
$n$ starting from 0. It is evident that an eigenfunction with $n<5$ depends
on $\tau_1$ only and its $\gamma$ is equal to zero. The eigenfunctions with
$4<n<8$ depend on $\tau_{1,2}$ only while the dependence on the $\tau_{1,2,3}$
occurs for the eigenfunctions with $7<n<12$. The eigenfunctions with $n \geq 12$
depend on all four variables $\tau_{1,2,3,4}$.

The function $\phi_{n,i}$ is related to the
eigenfunction of the Hamiltonian $\mathcal{H}_{H_4}$ (\ref{H_H4})
(and the integral ${\cal F}$) through
$\Psi_{n,i}=\Psi_{0}\phi_{n,i}$. Thus, the eigenfunctions $\{\phi\}$
are orthogonal with the weight factor $|\Psi_0|^2$. As an
illustration let us give explicit expressions for several
eigenfunctions $\phi_{n,i}$ and their respectful eigenvalues,

\begin{itemize}
\item $n=0$
\[
    \phi_{0,0}= 1\ ,\quad \ep_{0,0}= 0\ ,
\]
\item $n=1$
\[
    \phi_{1,0}= \om\tau_1 - 2(1+30\nu)\ ,\quad \epsilon_{1,0}= 2\om\ ,
\]
\item $n=2$
\[
  \phi_{2,0}= \om^2 \tau_1^2- 6\om (1+20\nu)\tau_1+ 6(1+20\nu)(1+30\nu)\ ,
  \quad \epsilon_{2,0}= 4\om\ ,
\]
\item $n=3$
\begin{equation*}
  \phi_{3,0}\ =\ \om^3 \tau_1^3- 12\om^2(1+15\nu)\tau_1^2+36 \om(1+15\nu)(1+20\nu)\tau_1-24(1+15\nu)(1+20\nu)(1+30\nu)\,,
\end{equation*}
\[
  \ep_{3,0}\ =\ 6 \om\ ,
\]
\item $n=4$
\[
   \phi_{4,0}= L_4^{(1+60\nu)}(\om \tau_1)\ ,\ \ep_{4,0}= 8\om\ ,
\]
\item $n=5$
\[
   \phi_{5,0}= L_5^{(1+60\nu)}(\om \tau_1)\ ,\ \ep_{5,0}= 10\om\ ,
\]
\[
     \phi_{5,1}\ =\ \om^6 \tau_2 - 3 (1+10\nu)\om^5 \tau_1^5 + 45\om^4(1+10\nu)^2 \tau_1^4 -
      300\om^3 (1+12\nu)(1+10\nu)^2 \tau_1^4 +
\]
\[
     900\om^2 (1+15\nu)(1+12\nu)(1+10\nu)^2 \tau_1^2 -
     1080\om (1+20\nu)(1+15\nu)(1+12\nu)(1+10\nu)^2 \tau_1 +
\]
\[
     360 (1+30\nu)(1+20\nu)(1+15\nu)(1+12\nu)(1+10\nu)^2\ ,
\]
\[
     \ep_{5,1} =12\om\ .
\]
\end{itemize}

Let us denote $\tilde \phi_{n,i}$ the common eigenfunctions of $h_{H_4}$
and $f$ which are elements of the invariant space
$P^{(1,6,10,15)}_n$ and their respectful eigenvalues $\tilde\ep_{n,i},
\gamma_{n,i}$. The index $i$ numerates these eigenfunctions for given
$n$ starting from 0. It is evident that an eigenfunction with $n<6$ depends
on $\tau_1$ only and its $\gamma$ is equal to zero. The eigenfunctions with
$6\leq n<10$ can depend on $\tau_{1,2}$ and the dependence on the $\tau_{1,2,3}$
occurs for the eigenfunctions with $10\leq n<15$. It is evident that all eigenstates $(n,i)$
at fixed $n$ and different $i$ are degenerate: their eigenvalues are equal to $2\om n$.
We give some eigenstates from $P^{(1,6,10,15)}_n$ explicitly,
\begin{itemize}
\item $n=0$
\[
    \tilde\phi_{0,0}= 1\ ,\quad \tilde\ep_{0,0}= 0\ ,\quad \gamma_{0,0}= 0\ .
\]
\item $n=1$
\[
    \tilde\phi_{1,0}= \om\tau_1 - 2(1+30\nu)\ ,\quad \tilde\ep_{1,0}= 2\om\ ,
    \quad \gamma_{1,0}= 0
\]
\item $n=2$
\[
  \tilde\phi_{2,0}= \om^2 \tau_1^2- 6\om (1+20\nu)\tau_1+ 6(1+20\nu)(1+30\nu)\ ,
  \quad \tilde\ep_{2,0}= 4\om\ ,\quad \gamma_{2,0}= 0
\]
\item $n=3$
\begin{equation*}
  \tilde \phi_{3,0}\ =\ \om^3 \tau_1^3- 12\om^2(1+15\nu)\tau_1^2+36 \om(1+15\nu)(1+20\nu)\tau_1-24(1+15\nu)(1+20\nu)(1+30\nu)\,,
\end{equation*}
\[
  \tilde\ep_{3,0} =6 \om\ ,\ \gamma_{3,0}= 0
\]
\item $n=4$
\[
   \tilde\phi_{4,0}= L_4^{(1+60\nu)}(\om \tau_1)\ ,\ \tilde\ep_{4,0}= 8\om\ ,
   \ \gamma_{4,0}= 0
\]
\item $n=5$
\[
   \tilde\phi_{5,0}= L_5^{(1+60\nu)}(\om \tau_1)\ ,\ \tilde\ep_{5,0}= 10\om\ ,\ \gamma_{5,0}= 0
\]
\item $n=6$
\[
   \tilde\phi_{6,0}= L_6^{(1+60\nu)}(\om \tau_1)\ ,\ \tilde\ep_{6,0}= 12\om\ ,\ \gamma_{6,0}= 0
\]
\[
   \tilde \phi_{6,1}\ =\ \tau_2-\frac{1+10\nu}{4(7+60\nu)}\tau_1^6\ ,\ 
     \tilde\ep_{6,1} =12\om\ ,\ \gamma_{6,1}= 12(7+60\nu)\ .
\]
\end{itemize}
It is worth noting that $\phi_{5,1}=\om^6 \tilde \phi_{6,1} + A \tilde\phi_{6,0}$
where $A$ is a parameter. Eigenfunctions $\phi_{n,0}\ =\ \tilde \phi_{n,0}$ at
$n \leq 6$.

\section{Conclusions}

We have shown that the $H_4$ rational system related to the
non-crystallographic root system $H_4$ is exactly solvable with the
characteristic vector $(1,5,8,12)$. This work complements the previous
studies of the rational (and trigonometric) models, related with
crystallographic root systems (e.g. \cite{Ruhl:1995} -
\cite{Turbiner:2001}, \cite{Turbiner:2005_1}) and non-crystallographic
root systems $I_2(k)$ \cite{TTW:2009} and $H_3$ \cite{Garcia:2010}.
A certain significance of exploration of the
$H_4$ rational system is due to a fact that this model is defined in
four-dimensional Euclidian space. There are very few
known exactly-solvable systems in this space --
five-body Calogero-Sutherland $(A_4)$ and $BC_4$
rational-trigonometric models among them. All of them
are completely-integrable.

Taking Coxeter invariants of $H_4$ as coordinates provided us a way
to reduce the rational $H_4$ Hamiltonian to algebraic form. It gave
us a chance to find the eigenfunctions of the rational $H_4$
Hamiltonian which are proportional to polynomials in these invariant
coordinates. It seems correct that these eigenfunctions exhaust all
eigenfunctions in the Hilbert space. It is worth noting that the
matrix $A_{ij}(\tau)$ which appears in front of the second
derivatives after changing variables in Laplacian from Cartesian to
the $H_4$ Coxeter invariant coordinates (see Eqs. (\ref{A-B})) has
polynomial entries corresponding to flat space metric, hence the
Riemann tensor vanishes.

It should be stressed that it was stated in Lax pair formalism
that the Hamiltonian of the $H_4$ rational system (\ref{H_H4})
is completely integrable \cite{Sasaki:2000}. This implies the
existence of three mutually-commuted operators (the `higher Hamiltonians')
which commute with the Hamiltonian forming a commutative algebra. It is
known (see \cite{Olshanetsky:1983}) for the crystallographic systems
that these higher Hamiltonians are the differential operators of the
degrees which coincide to the minimal degrees of the root space (the
Lie algebra) or their doubles for the $A_N$ case. It may suggest that
for the $H_4$ rational system the commuting integrals might be
differential operators of the orders 12, 20 and 30. Their explicit forms
are not known so far. It seems evident that these commuting operators
should take on an algebraic form after a gauge rotation (with the
ground state function as a gauge factor), and a change of variables
from Cartesian coordinates to the Coxeter invariant variables $\tau$'s.
Interesting open question is about a flag of invariant subspaces: would 
it be one of these two flags preserved by $h_{H_4}$?
Following the experience with different integrable systems, it seems the
integral(s) related with separation of variables do not enter to the
commutative algebra. Therefore, the integral ${\cal F}$ lays out of
the commutative algebra of integrals. It might serve as an
indication to a superintegrability of the $H_4$ rational system.

It should be pointed out that unlike the rational models of the 
crystallographic root spaces it is not possible to construct integrable 
(and exactly-solvable) trigonometric systems related to the 
non-crystallographic root spaces as a natural generalization of the 
Hamiltonian Reduction Method \cite{Olshanetsky:1983}.

The existence of algebraic form of the $H_4$ rational
Olshanetsky-Perelomov Hamiltonian makes possible the study of their
polynomial perturbations which are invariant wrt the $H_4$ Coxeter
group by purely algebraic means: one can develop a perturbation
theory in which all corrections are found by linear algebra methods
\cite{Tur-pert}. In particular, it gives a chance to calculate the $H_4$
Coxeter-invariant, polynomial correlation functions by algebraic means.

Another important property of the existence of algebraic form of the
$H_4$ rational Hamiltonian is a chance to perform a canonical,
Lie-algebraic discretization to uniform (\cite{ST:1995}) and exponential
\cite{Chriss:2001} lattices, or mixed uniform-exponential lattices. 
In the case of all three lattices such a discretization preserves a property 
of integrability, polynomiality of the eigenfunctions remains and 
it is isospectral.
Making the weighted projective transformation (\ref{wpt}) of the
$H_4$ algebraic form (\ref{h_H4_tau}) we arrive at different
algebraic form of the $H_4$ Hamiltonian. Making then the
Lie-algebraic discretization we arrive at a discrete model related
to an original discrete model via change of variables. It can be
considered as a definition of a polynomial change of variables for
discrete operators.

One can find the $sl(2)$-quasi-exactly-solvable generalization \cite{Turbiner:1988}
of the $H_4$ model which remains integrable. This is one of the first examples of
quasi-exact-solvability related to non-crystallographic root
systems. It complements the results obtained previously for all
rational models related to crystallographic systems (see
\cite{Turbiner:2005_2}), for the $I_2(k)$ rational model
\cite{TTW:2009} and for $H_3$ \cite{Garcia:2010} -- each of these models admit
a certain $sl(2)$-QES generalization in a form of sixth degree polynomial
potential.

Owing to the explicit knowledge of the ground state function (\ref{Psi_H4})
supersymmetric $H_4$ model can be constructed following a procedure
realized in \cite{Freedman:1990} for $A_N$ rational model, in
\cite{Brink:1998} for the $BC_N$ rational model and in \cite{Quesne:2010}
for the $I_2(k)$ rational model. It can be done elsewhere.

\bigskip

\textit{\small Acknowledgements}. The computations in this paper
were performed on MAPLE 8 and MAPLE 13 with the packages COXETER and
WEYL created by J.~Stembridge. The research is supported in part
by DGAPA grant IN115709 (Mexico). A.V.T. thanks the University Program
FENOMEC (UNAM, Mexico) for partial support.

\end{document}